\documentclass[letterpaper,twocolumn,prx,aps,superscriptaddress,floatfix]{revtex4}
 \usepackage[utf8]{inputenc}
\usepackage{bm}
\usepackage{amssymb,amsbsy,amsmath}
\usepackage{multirow,amssymb,amsbsy,amsmath,amsthm}
\usepackage{graphicx}
\usepackage{verbatim}
\usepackage{color}
\usepackage{pifont}
\usepackage{hyperref}
\usepackage{xcolor}
\usepackage{float}
\usepackage[absolute]{textpos}

\theoremstyle{theorem}

\theoremstyle{definition}

\newtheorem{definition}{Definition}

\theoremstyle{remark}

\hypersetup{
    colorlinks,
    linkcolor={blue},
    citecolor={blue},
    urlcolor={blue}
}

\makeatletter

\begin{document}

\title{Shedding Light on the Future: Exploring Quantum Neural Networks through Optics}

\author{Shang Yu}
%\altaffiliation{These authors contributed equally to this work}
\email{shang.yu@imperial.ac.uk}
\affiliation{Blackett Laboratory, Department of Physics, Imperial College London, Prince Consort Rd, London, SW7 2AZ, United Kingdom}
\affiliation{Centre for Quantum Engineering, Science and Technology (QuEST), Imperial College London, Prince Consort Rd, London, SW7 2AZ, United Kingdom}

\author{Zhian Jia}
%\altaffiliation{These authors contributed equally to this work}
\email{giannjia@foxmail.com}
\affiliation{Centre for Quantum Technologies, National University of Singapore, Queenstown 117543, Singapore}
\affiliation{Department of Physics, National University of Singapore, Queenstown 117543, Singapore}

\author{Aonan Zhang}
\affiliation{Blackett Laboratory, Department of Physics, Imperial College London, Prince Consort Rd, London, SW7 2AZ, United Kingdom}
\affiliation{Centre for Quantum Engineering, Science and Technology (QuEST), Imperial College London, Prince Consort Rd, London, SW7 2AZ, United Kingdom}

\author{Ewan Mer}
\affiliation{Blackett Laboratory, Department of Physics, Imperial College London, Prince Consort Rd, London, SW7 2AZ, United Kingdom}
\affiliation{Centre for Quantum Engineering, Science and Technology (QuEST), Imperial College London, Prince Consort Rd, London, SW7 2AZ, United Kingdom}

\author{Zhenghao Li}
\affiliation{Blackett Laboratory, Department of Physics, Imperial College London, Prince Consort Rd, London, SW7 2AZ, United Kingdom}
\affiliation{Centre for Quantum Engineering, Science and Technology (QuEST), Imperial College London, Prince Consort Rd, London, SW7 2AZ, United Kingdom}

\author{Valerio Crescimanna}
\affiliation{Blackett Laboratory, Department of Physics, Imperial College London, Prince Consort Rd, London, SW7 2AZ, United Kingdom}
\affiliation{National Research Council of Canada, 100 Sussex Drive, Ottawa, Ontario K1N 5A2, Canada}
\affiliation{Department of Physics, University of Ottawa, 25 Templeton Street, Ottawa, Ontario K1N 6N5 Canada}

\author{Kuan-Cheng Chen}
%\affiliation{Department of Materials, Imperial College London, Prince Consort Rd, London, SW7 2AZ, United Kingdom}
\affiliation{Centre for Quantum Engineering, Science and Technology (QuEST), Imperial College London, Prince Consort Rd, London, SW7 2AZ, United Kingdom}
\affiliation{Department of Materials, Imperial College London, Prince Consort Rd, London, SW7 2AZ, United Kingdom}

\author{Raj B. Patel}
\email{raj.patel1@imperial.ac.uk}
\affiliation{Blackett Laboratory, Department of Physics, Imperial College London, Prince Consort Rd, London, SW7 2AZ, United Kingdom}
\affiliation{Centre for Quantum Engineering, Science and Technology (QuEST), Imperial College London, Prince Consort Rd, London, SW7 2AZ, United Kingdom}

\author{Ian A. Walmsley}
\affiliation{Blackett Laboratory, Department of Physics, Imperial College London, Prince Consort Rd, London, SW7 2AZ, United Kingdom}
\affiliation{Centre for Quantum Engineering, Science and Technology (QuEST), Imperial College London, Prince Consort Rd, London, SW7 2AZ, United Kingdom}

\author{Dagomir Kaszlikowski}
\affiliation{Centre for Quantum Technologies, National University of Singapore, Queenstown 117543, Singapore}
\affiliation{Department of Physics, National University of Singapore, Queenstown 117543, Singapore}

\begin{abstract}
At the dynamic nexus of artificial intelligence and quantum technology, quantum neural networks (QNNs) play an important role as an emerging technology in the rapidly developing field of quantum machine learning. This development is set to revolutionize the applications of quantum computing. This article reviews the concept of QNNs and their physical realizations, particularly implementations based on quantum optics . We first examine the integration of quantum principles with classical neural network architectures to create QNNs. Some specific examples, such as the quantum perceptron, quantum convolutional neural networks, and quantum Boltzmann machines are discussed. Subsequently, we analyze the feasibility of implementing QNNs through photonics. The key challenge here lies in achieving the required non-linear gates, and measurement-induced approaches, among others, seem promising. To unlock the computational potential of QNNs, addressing the challenge of scaling their complexity through quantum optics is crucial. Progress in controlling quantum states of light is continuously advancing the field. Additionally, we have discovered that different QNN architectures can be unified through non-Gaussian operations. This insight will aid in better understanding and developing more complex QNN circuits.
\end{abstract}

\maketitle
\date{\today}

\section{Introduction}
Quantum computing is making significant progress toward machines that can achieve an advantage over classical computers, and new algorithms and applications that can benefit from this rapidly advancing domain are emerging continually. Applications for certain complex problems harness the principles of quantum mechanics to perform calculations with an efficiency unattainable by conventional computers. Technologies being pursued to build scalable quantum computers include superconducting qubits \cite{Arute2019quan}, trapped ions \cite{Schafer2018ions1,Guo2024ion2}, photons \cite{PSIQ2024} and neutral atoms \cite{Bluvstein2024log}.

A particularly promising platform for quantum computers uses photonics. Photons are less susceptible to decoherence and thermal noise than atomic or solid-state materials, making them ideal for transmitting quantum information over long distances. For example, the recently-developed photonic quantum machines \cite{Zhong2020quan,Bartolucci2023Fus,Yu2022univ,Madsen2022Q} are capable of executing some specific quantum algorithms at room temperature, a notable departure from the cryogenic environments required for other qubit types. The platform also benefits from a mature silicon photonics technology, meaning that large-scale quantum devices can be achieved by silicon or silicon nitride integrated photonic chip techniques \cite{Madsen2022Q,Wang2023Int}.
To achieve a fully-scalable universal quantum computer requires quantum error correction and fault tolerance are essential \cite{Sivak2023real,Ni2023bea}. Error-correction architectures, such as the surface code \cite{Krinner2022Rea}, are being developed to enable scalable processors. In the framework of photonic quantum computing, for example, the Gottesman-Kitaev-Preskill (GKP) state has been prepared \cite{Konno2024Log}, just one example of a protected logical qubit suitable for error correction.
These developments highlight the versatility and potential scalability of photonic approaches, potentially accelerating the timeline for practical quantum computing applications.

Recently, the ongoing development of quantum computing technology \cite{preskill2018quantum,Bharti2022noisy} has opened some new approaches for quantum machine learning to tackle computational challenges related to quantum data \cite{schuld2015introduction,biamonte2017quantum}.
Variational quantum algorithms (VQAs) \cite{peruzzo2014variational,khatri2019quantum,arrasmith2019variational,cerezo2020variational,cerezo2021variational} and, more specifically, quantum neural networks (QNNs) \cite{Abbas2021power,Killoran2019con,cong2019quantum,farhi2018classification,Pesah2021absence,liu2022rep,liu2023analytic}, stand out as some of the most promising applications in this area.
Meanwhile, both of these techniques rely on the concept of parameterized quantum circuit (PQC), a type of quantum circuit architecture equipped with adjustable parameters, such as the angle of rotation gates and phase shifters.

QNNs combine the techniques of quantum mechanics with the structure and function of neural networks, and are able to represent and process data (including quantum data) in ways that classical systems cannot efficiently achieve. By utilizing PQCs, QNNs can be trained and optimized for specific tasks, opening doors for advanced applications in quantum machine learning and potentially accelerating the practical using of quantum computing.

Traditional neural networks \cite{haykin2004comprehensive,haykin2009neural,lecun2015deep,abiodun2018state}, which are fundamental to modern artificial intelligence, have driven substantial advancements across numerous fields. For applications in physics, see, for example, Refs.~\cite{guest2018deep,Carleo2019machine,jia2019quantum}. However, these advancements are inevitably constrained by the limitations of classical computational systems. As we continue to generate and grapple with increasingly complex data and problems, the demand for more potent computational solutions is intensifying.

Quantum computing harnesses the laws of quantum mechanics for information processing and has the potential to solve certain problems intractable for classical computing systems~\cite{preskill1998,Nielsen2010, Schor-1997-factoring}. 
Specialized photonic quantum computers without error-correction capabilities have demonstrated a quantum-classical separation in computing power when compared to even the best classical algorithms~\cite{Madsen2022Q,Zhong2020quan}, which may find practical application in areas such as quantum chemistry, graph theory and drug discovery~\cite{Huh2015boso,Banchi2020mole,Sempere2022expe,Yu2022univ}.

The superposition of correlated states yields entanglement between subsystems, and this phenomenon imbues QNNs with unmatched speed~\cite{Sharma2022train} and capacity~\cite{Abbas2021power}.
As a cornerstone of quantum machine learning (QML)~\cite{Cerezo2022challenges}, which leverages quantum systems to significantly accelerate the processing and analysis of large, complex datasets, the potential applications of QNNs span numerous fields, from pattern recognition to cancer prediction~\cite{Jeswal2019Rec}, promising to deliver disruptive breakthroughs.

The transition of QNN from theoretical concepts to practical applications presents significant challenges, particularly due to existing losses and errors. As a predecessor to quantum optical neural networks (QONN), optical neural networks (ONNs)~\cite{Sui2020onnreview,Denz2013onn} have been extensively utilized in the training and inference of deep neural networks for real-time processing and scenarios requiring rapid data throughput~\cite{zhang2021optical,Xu202111tops,Wang2023Image,Chen2023deep,Zhu2022space}, owing to the significant noise resistance of classical light. From this perspective, the development of QONN is not starting from scratch. Building on the foundation of existing optical computing technologies, particularly their advancements in hardware, QONN can evolve within a relatively mature technological environment~\cite{steinbrecher2019quantum,Killoran2019con,Yamamoto2020cohe,Ian2023Light}.

In this perspective paper, we review several definitions of QNNs proposed in recent years, along with their applications in practical problems. Subsequently, we analyze the potential implementation of QNNs through quantum optics technology. This article also explores future prospects and challenges in these rapidly evolving fields. Given the anticipated quantum leap in computational power, understanding these topics is crucial for fully harnessing and directing the power of quantum computing.

\section{Quantum neural networks}

Classical neural networks \cite{haykin2004comprehensive,haykin2009neural,lecun2015deep} play a crucial role in machine learning applications.
Various classical neural networks have been introduced over the years. These include feedforward networks such as the perceptron and convolutional neural networks, as well as recurrent neural networks such as the Boltzmann machine (also known as the Hopfield network) and long short-term memory networks.
Similarly, QNN can be categorized based on their structure and functionality. These networks integrate classical neural network architectures with principles of quantum computation.
In this section, we provide an overview of the fundamental concepts associated with QNNs and present illustrative examples for clarity.

\subsection{Basic concepts of quantum neural network}

\begin{figure*}[hbt]
\centering
\includegraphics[width=0.81\textwidth]{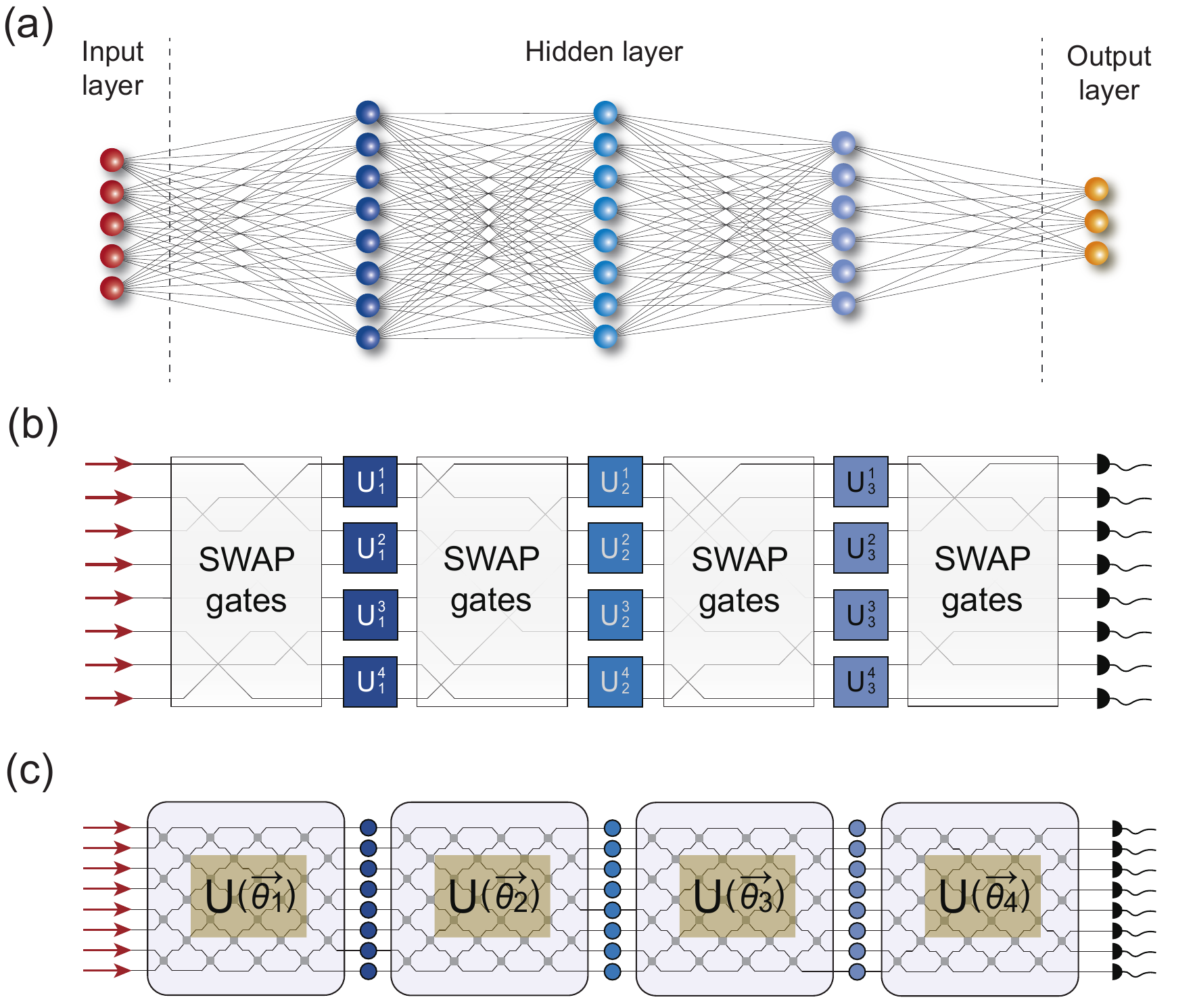}
\caption{\textbf{(a) Classical feedforward neural network and (b), (c) two QNNs with different definitions.} Under the definition shown in (b), QNN requires a quantum processor that can act on $n+1$ qubits, where the data is encoded in the first $n$ qubits, and the last qubit serves as a readout~\cite{farhi2018classification}. Here, the SWAP gate denotes a logic gate in quantum computing that swaps the states of two qubits. With the definition shown in (c),  the data is encoded in single photon Fock states, and a series of non-linear gates, for example, Kerr gates, implemented after the unitary operations~\cite{steinbrecher2019quantum}. The photon number at each output mode then is readout by photon-number-resolving detectors.}
\label{Fig1}
\end{figure*}

The foundational component of a neural network is the artificial neuron. This neuron receives several inputs, designated as $x_1,\cdots,x_n$, and generates a single output, $y$. Consider the feedforward neural network for example. Each input is associated with a specific weight, $w_i$. When the neuron receives the weighted sum $\sum_i w_i x_i$, it compares this sum with a bias, $b$, and applies an activation function, $f$, to the difference to yield an output value, ~\cite{mcculloch1943logical,rosenblatt1957perceptron,minsky2017perceptrons}:
$y=f(\sum_i w_i x_i-b)$.
A neural network is a complex system composed of numerous interconnected artificial neurons, structured according to a specified network architecture. See Fig.~\ref{Fig1} (a) for an illustration. These connections enable the  flow of information and computations that underpin the capabilities of the neural network.

A QNN synergizes classical neural network principles with quantum computational elements. Typically, a QNN is structured into three main components: (i) data encoding, which translates classical data into quantum states; (ii) a quantum circuit equipped with adjustable parameters that facilitate quantum computations; and (iii) quantum measurements, which extract relevant classical information from the quantum states.

The data is encoded in the quantum state, and the manipulation of these data is executed through quantum operations, such as unitary gates, quantum channels, or quantum measurements.
The encoding of classical data into a quantum state is not unique, for which there exist several proposals \cite{Wiebe2012quantum,schuld2017implementing}.

\begin{definition}[Data encoding]
The classical data $\mathcal{X}\subset \{0,1\}^n$ (or $\mathcal{X}\subset \mathbb{R}^n$) can be encoded in quantum states via a bijective map
\begin{equation}
    \vec{x}\mapsto \psi_{\vec{x}}, \forall \vec{x}\in \mathcal{X},
\end{equation}
where $\psi_{\vec{x}}$ is a pure state. This type of encoding is also called pure-state encoding. Similarly, we can introduce mixed-state encoding.
The data encoding is realized via a state preparation
circuit $U_{\vec{x}}$ acting on the initial state $|0\rangle^{\otimes n}$. For example, we have
\begin{enumerate}
    \item Basis encoding. For $n$-bit string $x\in \mathcal{X}$, choose a $n$-qubit Hilbert space $\mathcal{H}=(\mathbb{C}^{2})^{\otimes n}$ and maps $\vec{x}=(x_1,x_2\cdots ,x_n)\in \mathcal{X}$ to the basis 
    \begin{equation}\label{eq:angleencoding}
       \vec{x}\mapsto |\vec{x}\rangle=\bigotimes_{i=1}^n (\cos(x_i)|0\rangle + \sin(x_i)|1\rangle).
    \end{equation}
    \item Amplitude encoding. By introducing a feature map $\vec{f}:\mathcal{X}\to \mathbb{R}^N$, we can encode classical data in a $N$-dimensional feature Hilbert space
    \begin{equation}
        \vec{x}\mapsto |\psi_{\vec{x}}\rangle=\frac{1}{\|\vec{f}(\vec{x})\|_2} \sum_i f_i (x) |i\rangle,
    \end{equation}
    where $\|f\|_2=(\sum_i f_i(\vec{x})^2)^{1/2}$ and $i=1,\cdots,N$.
    A frequently used example of feature map $f$ is defined as $f_i(\vec{x})=x_i$, namely taking the $i$-th component of $\vec{x}$. In this case, the encoding is also called wavefunction encoding. 
    \item Angle encoding or product encoding. When $\vec{x}$ is an $n$-dimension real vector, the encoding given is Eq.~\eqref{eq:angleencoding} is also called angle encoding.
%    \item Phase encoding.
\end{enumerate}
\end{definition}

%Before we proceed, it's worth mentioning that data encoding is a crucial step in QNN and QML.

Unlike classical neural networks that utilize non-linear activation functions, as shon in Fig. 1(b), QNNs rely on PQCs as their foundational structure \cite{farhi2018classification}.

\begin{definition}[Parameterized quantum circuit (PQC)]
A parameterized quantum circuit is essentially a quantum circuit that incorporates gates with tunable parameters, often denoted as continuous variables. Specifically, consider two sets of gates: $\mathcal{G}_1=\{V_1,\cdots,V_n\}$, which contains gates without parameters, and $\mathcal{G}_2=\{U_1(\theta_1),\cdots,U_{m}(\theta_m)\}$, consisting of gates with adjustable parameters. A quantum circuit constructed using the combined gate set $\mathcal{G}=\mathcal{G}_1\cup \mathcal{G}_2$ is termed a parameterized quantum circuit.
\end{definition}
    
A QNN can be regarded as a PQC $U_{\rm QNN}(\vec{\theta})$ defined by a specific circuit architecture, where $\vec{\theta}$ represents the adjustable parameters  \cite{farhi2018classification}.
We train the QNN via using the training set of data $\mathsf{Train}=\{(\vec{x}_i,{y}_i)\}$ where ${y}_i$'s are the labels of data $\vec{x}_i$'s. After the training, the parameters of the QNN are fixed, our goal is that on the test data set $\mathsf{Test}=\{(\vec{s}_i,{t}_i)\}$, we could obtain the correct label of data via observing some obserbable $O$ (or equivalently, measuring the output state in some given bases):
\begin{equation}
  t_i = \langle \psi_{\vec{s}_i}|U_{\rm QNN}(\vec{\theta})^{\dagger} O U_{\rm QNN}(\vec{\theta})|\psi_{\vec{s}_i}\rangle.
\end{equation}
To simplify the discussion, hereinafter, we will assume that label of all data is a real number.

Given that the fundamental architecture of a QNN is a PQC, it is important to determine the appropriate choice of ansatz gates.
A commonly used ansatz is of the form (see, e.g.,  \cite{Pesah2021absence,liu2022rep,liu2023analytic})
\begin{equation}
    U(\vec{\theta})=\prod_{k}e^{-i\theta_k H_k}W_k,
\end{equation}
where $H_k$ are Hermitian operators, $W_k$ are unparametrized gate (such as CNOT gate), and $\theta_k$ are parameters.

The parameters of the QNN are tuned based on a loss function, which can vary based on the QNN's architecture and the specific tasks it addresses.
Suppose that $O$ is our target observable, we can express the loss function as
\begin{equation}
    L(\vec{\theta})=\frac{1}{|\mathcal{X}_{\rm train}|}\sum_{k: \mathsf{Train}} c_k (\operatorname{Tr}(\rho_{\rm out}^k(\vec{\theta}) O) -y_k)^2, 
\end{equation}
where $c_k$ are real coefficients and $\rho_{\rm out}^k(\vec{\theta})=U_{\rm QNN}(\vec{\theta})\rho_{in}^kU_{\rm QNN}(\vec{\theta})^{\dagger}$ with $\rho^k_{\rm in}\in \mathcal{X}_{\rm train}$ the training quantum data with labels $y_k$.

A crucial problem in studying QNN is the trainability.
For a substantial number of QNNs, the cost function gradients of an ansatz, when randomly initialized, experience an exponential decrease as the problem size expands. This widely observed occurrence is referred to as the \emph{barren plateau} phenomenon \cite{mcclean2018barren}.
We denote the partial derivative of loss function with respect to $\theta_{\alpha}$ as $\partial_{\alpha}L:=\partial L(\vec{\theta})/\partial \theta_{\alpha}$.
It was pointed out in Refs. \cite{mcclean2018barren,cerezo2021cost,Sharma2022train,cerezo2021higher} that the trainability of a randomly initialized QNN can be analyzed by studying the scaling of the variance
\begin{equation}
    \operatorname{Var}[\partial_{\alpha}L]=\langle (\partial_{\alpha}L)^2\rangle-\langle \partial_{\alpha}L\rangle^2,
\end{equation}
where the expectation value is taken over the parameters. Using the assumption that $\langle \partial_{\alpha}L\rangle=0$, from Chebyshev's inequality, we obtain that 
\begin{equation}
    \operatorname{Pr}[|\partial_{\alpha}L|>\varepsilon]\leq \frac{ \operatorname{Var}[\partial_{\alpha}L]}{\varepsilon^2},
\end{equation}
where $\varepsilon\geq 0$.
If the variance, $\operatorname{Var}[\partial_{\alpha}L]$, is exponentially small, it indicates a barren plateau in the loss function. In such scenarios, the gradient of the loss function becomes vanishingly small on average, necessitating an exponentially high precision to traverse this flat region effectively.

\subsection{Examples of quantum neural network}

In this section, we will delve into specific examples of QNNs. Remarkably, QNNs often lack nonlinear activation functions, a feature we'll delve into in the upcoming examples.

\vspace{1em}
\emph{Quantum perceptron.} ---
Classical perceptrons can be regarded as the fundamental building block of more complex artificial neural networks \cite{mcculloch1943logical,rosenblatt1957perceptron,minsky2017perceptrons}.
The quantum perceptron is the quantum generalization, where data and weights  are encoded into quantum states.
The main tool to implement the quantum perceptron is Grover's search algorithm \cite{kapoor2016quantum}.
For the labeled data set $\mathcal{X}=\{(\vec{x}_i,y_i)\}$ (for simplicity, we assume the label $y_i$'s take values in $\{0,1\}$) and a given perceptron, our goal is to find weights that correctly classify the data.
This is characterized by a perceptron classification function: $F(\vec{x}_i,y_i;\{w_k\})=1$ if and only if the perceptron with weights $\{w_i\}$ misclassifies the data.
In the classical implementation of the online perceptron, training points are sequentially fed into the classification algorithm. Each time a point is misclassified, the weight parameters are updated.
The quantum version of the online perceptron diverges from the traditional sequential data access during an epoch. It employs a method of accessing data points in a superposed quantum state and applies the classification function to this state. This enables all data points to be processed simultaneously, streamlining the search for the misclassified point.
First, we use a quantum circuit to prepare the superposition of the data
 $  U_{\mathcal{X}}:|j\rangle \otimes |0\rangle \mapsto |j\rangle \otimes |x_i\rangle$.
Then we have
\begin{equation}
    U (\frac{1}{\sqrt{|\mathcal{X}|}}\sum_j |j\rangle) \otimes |0\rangle =\frac{1}{\sqrt{|\mathcal{X}|}} \sum_j |j\rangle \otimes |x_j\rangle.
\end{equation}
To implement Grover's search, we need to build a quantum circuit that implements the Boolean function $F(\vec{x}_i,y_i;\{w_k\})$.
With access to such a circuit, we can subsequently define a corresponding oracle quantum operator
\begin{equation}
    U_{F(\vec{x}_i,y_i;\{w_k\})} |x_j\rangle =(-1)^{F(\vec{x}_i,y_i;\{w_k\})} |x_j\rangle.
\end{equation}
Notice that $ U_{F(\vec{x}_i,y_i;\{w_k\})} $ depends on the weights $\{w_k\}$.
Now we can use $ U_{F(\vec{x}_i,y_i;\{w_k\})} $ as an oracle to implement the Grover search.
The quantum perceptron algorithm works as follows: (i) Apply Grover's search using $ U_{\mathcal{X}}$ and $ U_{F(\vec{x}_i,y_i;\{w_k\})} $ over the state $\frac{1}{\sqrt{|\mathcal{X}|}}\sum_j |j\rangle) \otimes |0\rangle$; (ii) Measure the first register, if $F(\vec{x}_i,y_i;\{w_k\})=1$, then update the weights $\{w_k'\}\leftarrow \{w_k\}$ and update the operation  accordingly $U_{F(\vec{x}_i,y_i;\{w'_k\})} \leftarrow\ U_{F(\vec{x}_i,y_i;\{w_k\})} $; (iii) Repeat the preceding two steps until the condition $\operatorname{Pr}[\exists j, F(x_j,y_j;\{w_k\}))=1] \leq \varepsilon$ is satisfied for a given small  $\varepsilon >0$.
For alternative approaches to quantum perceptron implementation, refer to the detailed analysis in Ref. \cite{kapoor2016quantum}.

\begin{figure}[t]
\centering
\includegraphics[width=0.45\textwidth]{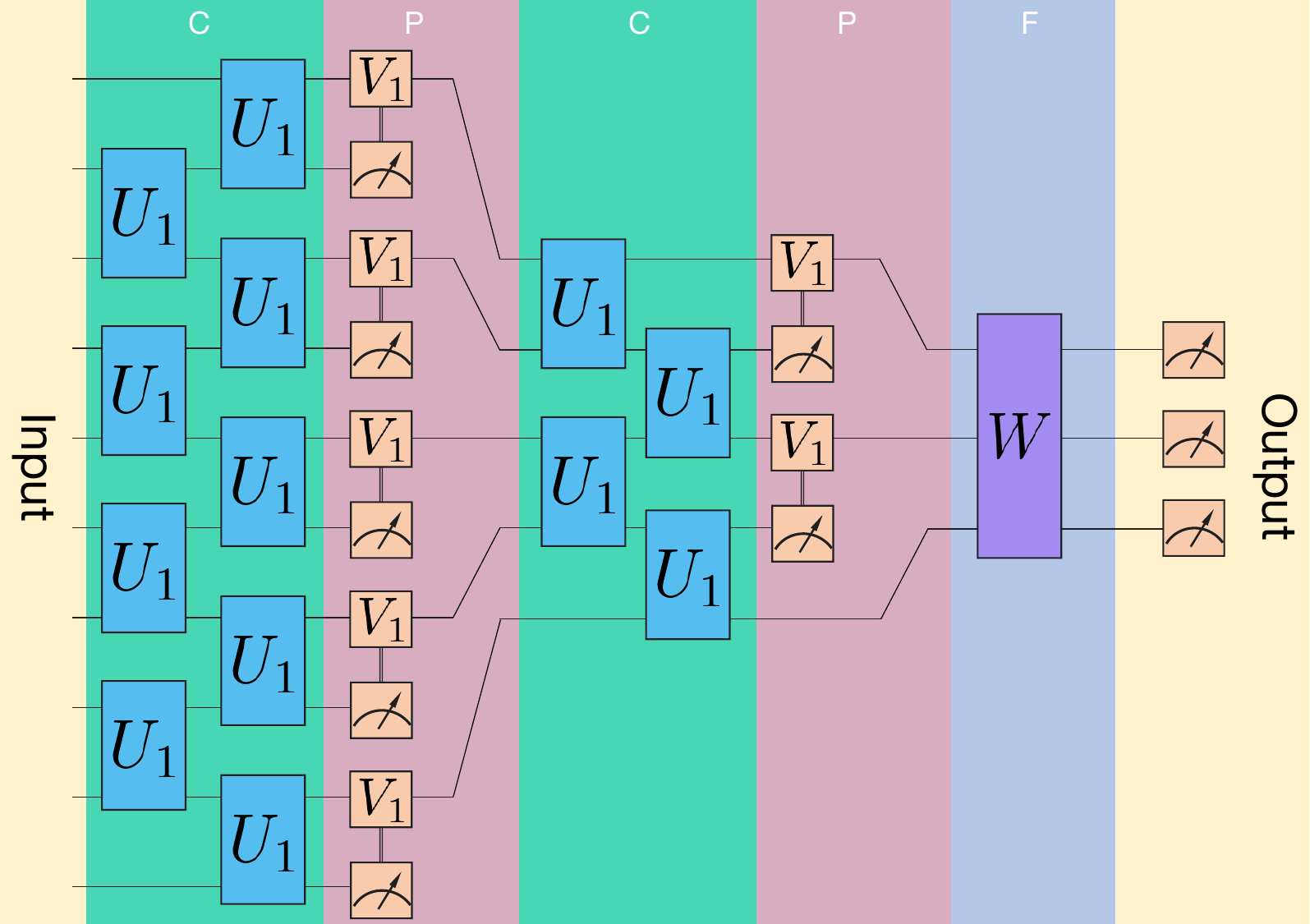}
\caption{\textbf{Depiction of a QCNN.} The input of the QCNN is an $n$-qubit state. In the convolutional layer, a local unitary gate $U_1$ is applied in a translationally invariant manner for a finite depth. In the pooling layer, certain qubits are measured, and the application of unitary gates $V_1$ is conditioned on the measurement outcomes. The fully connected layer involves the application of a global unitary gate $W$. Finally, at the last layer, the outcome state is measured.
}
\label{Fig_QCNN}
\end{figure}

\vspace{1em}
\emph{Quantum convolutional neural networks (QCNN)} ---
In classical computing, convolutional neural networks comprise layers for convolution, pooling, and full connectivity. Their quantum counterpart has been suggested in Ref. \cite{cong2019quantum}. 
See Fig.~\ref{Fig_QCNN} for an illustration.
For the convolution layer, we implement local unitary gate $U_1$ in a translationally invariant manner for finite depth. This is inspired the classical convolution operation, where a weighted convolutional kernel is applied in a translationally invariant manner.
In the pooling layer, a subset of qubits are measured. Subsequently, unitary gates $V_1$ are applied to the adjacent qubits, conditioned on the received measurement outcomes.
In the pooling layer, due to the reduction in the number of qubits, non-linearity is inherently introduced.
The fully connected layer is realized by applying a global unitary gate $W$.
For a  $n$-qubit input state $\rho_{\rm in}$, the output state $\rho_{\rm out}(\vec{\theta})$ has a much smaller dimension. 
The parameters, represented by $\vec{\theta}$ in the output state, require optimization. Typically, these parameters originate from the unitary gates $U_1$ in the convolutional layer and the global unitary gate $W$ in the fully connected layer.

In the study by Ref. \cite{Pesah2021absence}, it was demonstrated that barren plateaus are absent in QCNNs. This suggests that QCNNs can be effectively trained from random initializations, highlighting their potential in quantum data applications.

\vspace{1em}
\emph{Quantum Boltzmann machine.} ---
The classical Boltzmann machine is constructed using the energy function:
\begin{equation}
E=-\sum_{i,j}v_i w_{i,j} h_j -\sum_{i}a_i v_i -\sum_j b_j h_j,
\end{equation}
where $w_{i,j}$ are weights between visible neurons $v_i$ and hidden neurons $h_j$, $a_i$ and $b_j$ are biases.
The probability distribution is given by Boltzmann distribution
\begin{equation}
     \operatorname{Pr}(\{v_i\})=\frac{1}{Z}\sum_{\{h_j\}}e^{-E(\{v_i\},\{h_j\})},
\end{equation}
where $Z$ is the partition function.
The quantum Boltzmann machine is a quantum generalization\cite{Amin2018quantum,Rebentrost2018hopfield}. Following proposal of Ref. \cite{Amin2018quantum}, both the hidden and visible neurons are replaced with quantum spins and the energy function is replaced with a Hamiltonian:
\begin{equation}
    H=-\sum_i b_i \sigma_i^z-\sum_{i,j}w_{i,j}\sigma^z_i\sigma_j^z.
\end{equation}
The corresponding thermal equilibrium state is:
\begin{equation}
    \rho= \frac{e^{-H}}{Z},
\end{equation}
where $Z=\operatorname{Tr} e^{-H}$.
To obtain the Boltzmann probability of quantum visible neurons, we construct the operator 
\begin{equation}
    \Lambda_{\{v_i\}} =|\{v_i\}\rangle \langle \{v_i\}|\otimes I.
\end{equation}
The marginal Boltzmann probability is obtained by
\begin{equation}
    \operatorname{Pr}(\{v_i\}) =\operatorname{Tr}(\Lambda_{\{v_i\}} \rho).
\end{equation}
Consequently, we can conduct supervised learning analogously to the classical Boltzmann machine approach.

The above three examples are not exhaustive, as numerous quantum neural network models have been proposed. However, since this work focuses primarily on constructing quantum neural networks through optical methods, we will not delve further into this topic.

\section{Realizing quantum neural networks with optics}
Optical quantum computing possesses significant advantages, especially in the following key areas. Firstly, unlike other quantum systems that require low-temperature, it can operate in ambient environments. Secondly, photons have minimal interactions with the environment, giving these systems a high degree of noise isolation. Besides, optical quantum computing can be easily integrated with existing optical communication networks, offering substantial convenience for practical applications and system expansion.

However, no known or anticipated material possesses an optical nonlinearity strong enough to implement the deterministic photon-photon gate~\cite{OBrien2007opti}, which typically requires a strong light-matter coupling system~\cite{Hacker2016aphon}. As a result, quantum computation with optics also naturally bring a notable drawback: the weak interaction between photons makes the implementation of two-qubit quantum gates challenging and probabilistic, thereby limiting its scalability~\cite{Knill2001klm}.

\begin{figure*}[hbt]
\centering
\includegraphics[width=1.00\textwidth]{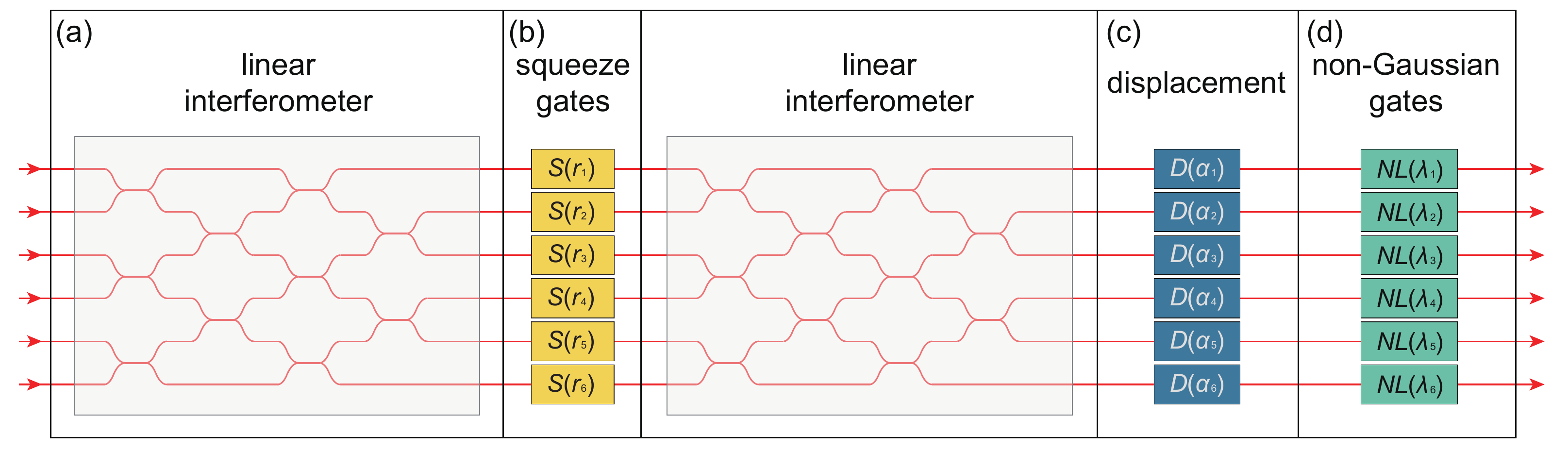}
\caption{\textbf{Realization the QONN in CV architecture~\cite{Killoran2019con}}. In CV architecture, the modules to realize QONN are mainly divided as four parts: (a) unitary operations achieved by linear operators, (b) measurement-based squeezing gate, (c) displacements with local oscillators, and (d) non-linear operation realized by non-Gaussian gates, which usually achieved by photon-additions and photon-subtractions (including PNRD) nowadays.}
\label{Fig3}
\end{figure*}

In order to realizing QONN under ccontinuous variable (CV) architecture (as shown in Fig. 1 (c)), linear transforms and non-linear gates are necessary~\cite{steinbrecher2019quantum,Killoran2019con}. While the former one is easy to achieve in the normal optical circuit~\cite{Yu2022univ,Arrazola2021Quan,Bao2023veryl}, but the non-linear gate (e.g., Kerr-type gate) is quite challenge to realize, especially in the weak field case. What's more, the lack of strong optical nonlinearity, as discussed above, also hinders the implementation of deterministic two-qubit entangling gates in the digital variable (DV) architecture~\cite{Knill2001klm}, which is crucial for constructing a universal photonic quantum computer.

We've discovered that many operations in optical systems are relatively more straightforward to implement within the CV architecture. For instance, it allows for the preparation of large-scale cluster states with distinct advantages~\cite{Larsen2019Deter,Asavanant2019Gener}, and achieving programmable Gaussian Boson sampling circuit~\cite{Zhong2021phas,Madsen2022Q,Yu2022univ}. What's more, besides the linear transformations and squeezing operations can be realized in these above experiments, notably, the realization of non-linear gates have also become more feasible in this architecture compared with directly introducing non-linear crystal in weak field, and there is hope for deterministic implementation~\cite{Costanzo2017Meas}.

Leveraging the convenience of implementing the aforementioned operations, let's first explore how to realize QONN in optical systems under the CV architecture~\cite{Killoran2019con}, the corresponding circuit is shown in Fig. 3.

As shown in Fig. 3(a), a linear interferometer need to be applied into the QONN at beginning. To accommodate the parameter adjustments required in QONN, the linear interferometer must be programmable at will. Such a linear interferometer can currently be easily implemented using integrated quantum photonic chips~\cite{Arrazola2021Quan} (as shown in Fig. 4(a)) and time-bin loop-based processor structures~\cite{Madsen2022Q,Yu2022univ,Yu2022von} (as shown in Fig. 4(b)). Both of these architectures can implement arbitrary unitary operations and are easily scalable in terms of mode numbers.

\begin{figure}[hbt]
\centering
\includegraphics[width=0.49\textwidth]{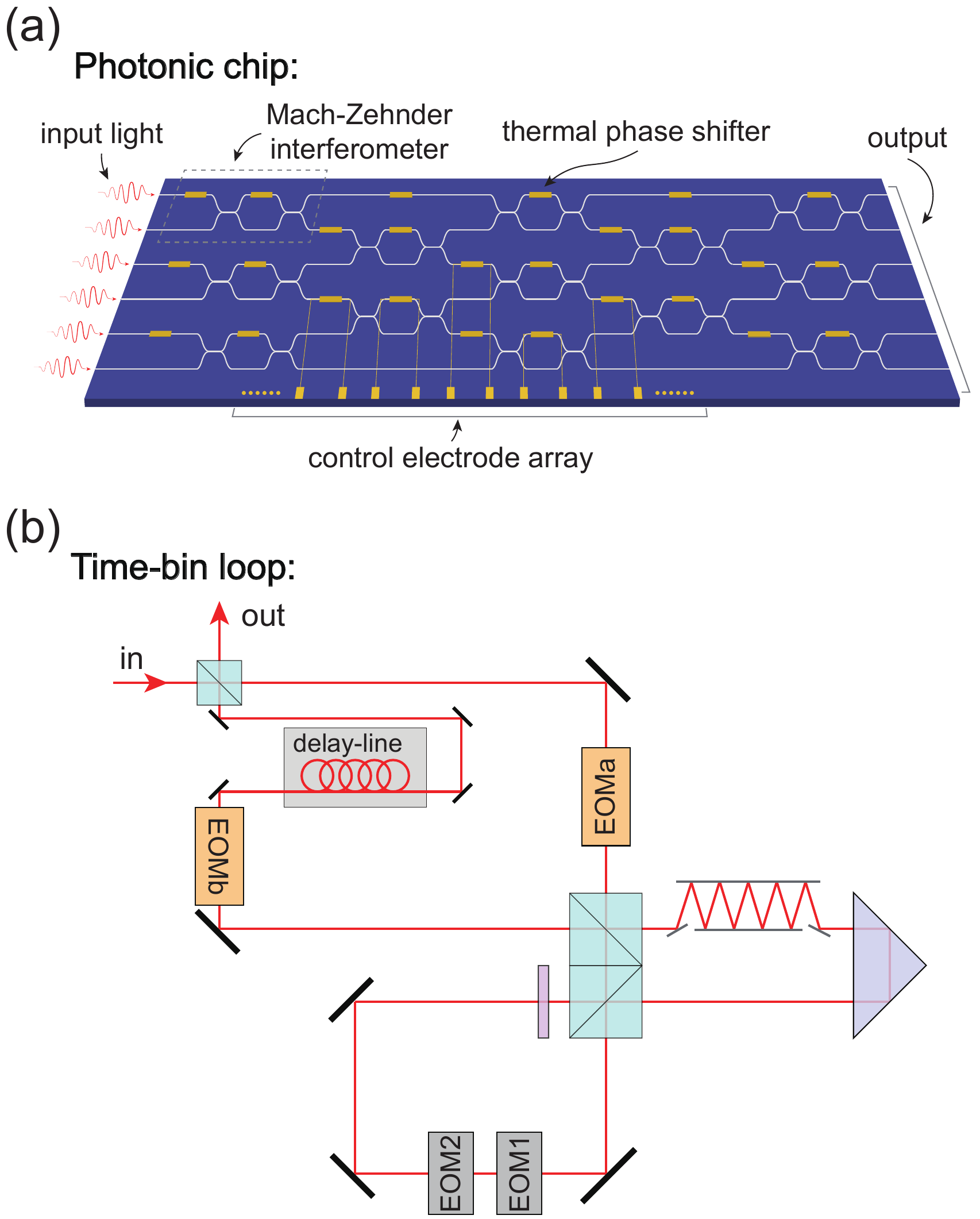}
\caption{\textbf{Unitary operation realized with all-optical devices}. (a) Integrated photonic chips~\cite{Arrazola2021Quan,Bao2023veryl} design according to Clements' decomposition~\cite{Clements2016opti}. (b) Unitary operation realized with time-bin loop setup~\cite{Yu2022univ,Yu2022von}.}
\label{Fig4}
\end{figure}

The optical quantum chip system involves using integrated photonic circuits (shown in Fig. 4(a)) to perform fully programmable unitary operations with high precision, leveraging the stability and low decoherence of photonic systems. Such an integrated chip system is scalable and can be fabricated using well-established manufacturing techniques, making them a promising option for building large-scale QONNs.

Fully programmable unitary operations can also be realized with time-bin loop setups, as shown in Fig. 4(b). This method uses time-bin encoding, whereby information is stored in every pulse of light, to manipulate qumodes. in this arrangement electro-optic modulators (EOMs) can effectively implement the unitary gate on the polarization degree of freedom \cite{Yu2022univ,Yu2022von}, creating a robust and flexible framework for photonic quantum computations. Both methods highlight the full programmability and stability needed to overcome the limitations of current quantum hardware.

Squeezing gates (as shown in Fig. 3(b)) are followed by linear interferometers. The effect of the squeezing gate on Fock basis states $x_{i}$ is
\begin{equation}
    \hat{S}(r_{i})|x_{i}\rangle=\sqrt{c_{i}}|c_{i}x_{i}\rangle
\end{equation}
where $c_{i} = e^{-r_{i}}$~\cite{Killoran2019con}, and $r_{i}$ denotes the squeezing level. This operation can be implemented using a measurement-induced squeezing gate with all optical system~\cite{Yoshikawa2007Demo,Takeda2017Uni}. In addition to providing an input, this operation also requires the simultaneous input of an auxiliary squeezed state. These two states are then both input into a polarizing beam splitter (PBS) with an adjustable transmission-reflection ratio. A displacement operation is needed to be performed based on a measurement of the field quadrature amplitude on the auxiliary output arm. The final output state then can be considered as applying a squeezing operation to the input state. The degree of squeezing in this operation is determined by the transmission-reflection ratio of the PBS~\cite{Yoshikawa2007Demo,Takeda2017Uni}. Assume transmission ratio of the PBS is $T_{0}$, then the $x_{\text{output}}$ quadrature of the output state becomes $x_{\text{output}}=\sqrt{T_{0}}x_{\text{input}}+\sqrt{1-T_{0}}x_{\text{ancilla}}e^{-r_{\text{ancilla}}}$~\cite{Yoshikawa2007Demo}. We can observe that this operation is equivalent to a squeezing operation of magnitude $r=-\ln{\sqrt{T_{0}}}$. Moreover, as the squeezing level of the auxiliary squeezed state increases, the resulting output state approaches the ideal state more closely.

The subsequent displacement operation (as shown in Fig.3(c)) is relatively straightforward. One simply needs to interfere the target state with a local oscillator on a beam splitter (BS) with a 99:1 ratio. We can place a phase modulator before the BS to adjust the phase of the input local oscillator~\cite{Thekkadath2022expe}. This can effectively implement the $D(\alpha)$ operation, and $\alpha=|\alpha|e^{-\varphi}$.

Lastly, we introduce the most crucial nonlinear operation as depicted in Fig. 3(d). Typically characterized as a non-Gaussian operation~\cite{Zhuang2018reso}, this function plays an essential role within the CV framework. Numerous operations fall under this category, such as photon number resolved detection (PNRD)~\cite{Namekata2010Non}, Kerr gate~\cite{Costanzo2017Meas}, and cubic phase gate~\cite{Miyata2016Impl}. The Kerr-type nonlinear operation, for example, has only been realized through a measurement-induced method, involving a process of photon addition and subtraction~\cite{Costanzo2017Meas}. The observation can be made that within the CV architecture, all requisite operations can be executed through exclusively optical devices. Drawing upon existing technologies, like integrated optical chips and free-space time-bin systems, the implementation of QONN demonstrates substantial scalability. This suggests that, in the future, we could train larger datasets with QONN and genuinely transition QONN into practical application.
Meanwhile, analyzing the nonlinear operation realized by this non-Gaussian process is also beneficial for preparing GKP states~\cite{Bourassa2021Blue,Takase2023gkp,Konno2024Log} and advancing the study of fault-tolerant quantum computers~\cite{Bourassa2021Blue}.

However, quantum hardware is currently still in its nascent stage, often referred to as the noisy intermediate-scale quantum computing (NISQ) era. The limited scale and relatively high error rates pose significant challenges for scaling QNNs to practical sizes. Thus, effective error correction techniques are essential for maintaining the integrity of computations.

Besides, optimization strategies and randomness play crucial roles in the training process. Optimization strategies include quantum natural gradient descent, hybrid classical-quantum optimization, and parameterized quantum circuits, which adjust parameters to minimize the loss function. Randomness in QNNs also manifests in random initialization, random sampling, and random perturbations, and can help to avoid local minimum and estimate gradients.

Additionally, we can integrate QONNs under both CV/DV-architectures with Gaussian and non-Gaussian operations~\cite{Bourassa2021fast}. In the CV-architecture (like shown in Fig.~\ref{Fig1} (c)), non-Gaussian operations are used to accomplish the highly significant and most challenging non-linear operations. Whereas in the DV-architecture (like shown in Fig.~\ref{Fig1} (b)), such non-Gaussian resources can potentially be used to complete two-qubit gate operations utilizing GKP states~\cite{Kyungjoo2022low}. Therefore, we can find that both CV and DV architectures require the intervention of non-linear/non-Gaussian operations to effectively implement QONN.

While the realization of QNNs using quantum optics still faces many challenges, particularly in terms of scalability and noise reduction, the theoretical underpinning and preliminary experimental results indicate a promising path forward. Continued advancements in the control and manipulation of quantum states of light are expected to drive progress in the development of practical QNNs.

\section{Potential applications of QNN}
One prominent application of QNNs is in the field of handwritten digit recognition \cite{Zhou1999Rec}. In this case, a QNN combines the advantages of neural modelling and fuzzy theory. It is designed to apply to both real data and synthetically distorted images, has has been shown to give both efficiency and accuracy enhancement in identifying handwritten numbers \cite{Zhou1999Rec}. With the continuous improvement of technology, QNN has potential for broader applications in pattern recognition \cite{Xu2011quan,Mu2013lea}. Moreover, QNNs hold significant promise in medical diagnostics, particularly in predicting breast cancer \cite{Li2013Mod}. This application underscores the transformative potential of QNNs in improving healthcare outcomes through advanced predictive analytics.

Furthermore, the potential applications of QNNs are vast and varied, impacting such fields as quantum phase recognition \cite{cong2019quantum}, artificial intelligence \cite{Altaisky2020Qua}, and weather prediction \cite{Safari2021Qua}. As research and development in quantum computing continue to advance, the deployment of QNNs could lead to breakthroughs that leverage quantum advantages for solving some complex problems with unprecedented efficiency and accuracy.

\section{Challenges and Future Directions}

While developing and implementing QNNs through quantum optics shows great promise, it also presents some unique challenges that must be overcome. Scalability is one of the significant technical difficulties that need to be addressed. As the scale of quantum systems increases, the complexity of managing the system also increases. This is particularly relevant to quantum neural networks, where the complexity will grow exponentially with the number of neurons (nodes) and synapses (connections).

As the scale of QNNs expands, quantum systems become prone to errors due to losses and operational defects. Then the effective quantum error-correction codes and techniques are essential for practical QONN applications, but these technologies are still underdeveloped.

The development of quantum neural networks and quantum optics has immense potential in shaping the future of computing, offering new approaches to current computational challenges. QNNs are predicted to facilitate the development of new quantum algorithms. These algorithms could significantly improve the efficiency of data processing tasks in various areas, from financial modeling to drug discovery, such as pattern recognition and optimization problems. Additionally, with the aid of QNNs, quantum machine learning is possible to revolutionize the field of artificial intelligence by providing more powerful models. This could lead to breakthroughs in complex tasks like natural language processing, image recognition, and real-time decision-making. Meanwhile, with advancements in quantum optics and integrated optical circuits, it is possible to develop scalable and practical quantum systems. As the technology matures, we can expect to see QNNs integrated into everyday computing devices, accelerating the advent of the quantum computing era. Furthermore, the integration of quantum neural networks in quantum optics might assist in establishing a quantum internet. This would enable ultra-secure communication, distributed quantum computing, and a new level of coordination between quantum devices.

While these opportunities are promising, they come with significant challenges that necessitate thorough research and development. Nonetheless, persistent effort and investment in quantum neural networks and quantum optics could significantly enhance our computational power.

In summary, with the quantum domain poised for computational leaps, the fusion of quantum neural networks and quantum optics is at the forefront of this evolution. While the journey to fully harness and steer the power of quantum computing remains challenging, the ongoing innovations and research detailed in this article set a hopeful trajectory for the quantum revolution.

\section*{Acknowledgements}
We thank Steven Sagona-Stophel for providing useful feedback on the manuscript. 
This work was supported by Imperial QuEST seed funding, UK Research and Innovation Guarantee Postdoctoral Fellowship (EP/Y029631/1), Engineering and Physical Sciences Research Council and Quantum Computing and Simulation Hub (T001062), UK Research and Innovation Future Leaders Fellowship (MR/W011794/1),, EU Horizon 2020 Marie Sklodowska-Curie Innovation Training Network (no. 956071, `AppQInfo'), National Research Foundation in Singapore, and A*STAR under its CQT Bridging Grant.

\bibliographystyle{apsrev4-1-title}
\bibliography{mybib}

\end{document}